%% file: main.tex
\documentclass[11pt]{article}
\pdfoutput=1
\usepackage[utf8]{inputenc}
\usepackage[left=1in,right=1in,top=1in,bottom=0.7in]{geometry}
\usepackage{microtype}
\usepackage{amsmath,amssymb}
\usepackage{graphicx}
\usepackage{booktabs,tabularx,array}
\usepackage{caption,subcaption}
\usepackage{setspace}
\usepackage{placeins}
\usepackage{xcolor}
\usepackage[round,authoryear]{natbib}
\usepackage{xurl}
\definecolor{linkblue}{RGB}{25,76,153}
\usepackage[colorlinks=true,citecolor=linkblue,linkcolor=linkblue,urlcolor=linkblue]{hyperref}
\hypersetup{pdftitle={AutoAnalysis: An automated, self-updating meta-analysis of human-AI collaboration},pdfauthor={Michelle Vaccaro}}
\renewenvironment{abstract}{%
  \par\begin{center}\bfseries Abstract\end{center}%
  \begingroup\setstretch{1.05}\setlength{\parindent}{0pt}\noindent\ignorespaces
}{\par\endgroup}

\title{\textbf{AutoAnalysis: An automated, self-updating meta-analysis of human-AI collaboration}}
\author{
Michelle Vaccaro\textsuperscript{1},
Abdullah Almaatouq\textsuperscript{2},
Thomas W. Malone\textsuperscript{2}\\[0.5em]
\normalsize \textsuperscript{1}Harvard Business School\\
\normalsize \textsuperscript{2}MIT Sloan School of Management
}

\date{October 26, 2026}

\begin{document}
% Keep title, author, date and abstract together on page 1.
% In particular, do not put a \newpage between \maketitle and the abstract.
\newgeometry{left=1in,right=1in,top=0.6in,bottom=0.7in}
\begingroup
\setstretch{1}
\maketitle
\endgroup
\vspace{-0.5em}
\begin{abstract}
As generative artificial intelligence (AI) becomes more capable and widely used, understanding when and how humans should work with these systems is becoming increasingly important. However, the empirical evidence in this space offers conflicting recommendations, which leaves individuals and organizations without clear guidance on when to combine human and AI capabilities. Existing meta-analyses can help reconcile these findings, but they require substantial manual effort and cannot readily incorporate new evidence as it emerges. To address this challenge, we introduce AutoAnalysis: an automated, self-updating meta-analysis of when humans and AI work better together, and when they work better alone. AutoAnalysis' most recent run synthesizes 1,955 effect sizes from 420 papers. Across this body of work, we found that, on average, human-AI combinations performed significantly better than humans alone, but they performed significantly worse than the better of humans or AI alone. These findings highlight the important distinction between improving human performance and achieving human-AI synergy and underscore the need to evaluate collaboration against both human and AI performance alone. As new studies emerge, AutoAnalysis will integrate their findings into the meta-analysis, providing cumulative evidence about when humans and AI perform better together and how human-AI performance varies across AI agents, tasks, and processes.

\end{abstract}

\vfill

\begingroup
\footnotesize
\singlespacing
\noindent Please visit
\href{https://auto-analysis.dev}{auto-analysis.dev} to explore the data and view the most recent results.

\noindent Generative AI disclosure: We used generative AI to help develop AutoAnalysis and to help write this manuscript. This use of AI is part of our broader research effort to understand how these tools can support social science research. We carefully reviewed the resulting work and take full responsibility for the paper's content and conclusions.
\par
\endgroup

\clearpage

\clearpage
\restoregeometry

\input{body}

% Flush figures before the bibliography; supplementary material follows it.
\clearpage
\begingroup
\singlespacing
\small
\phantomsection
\addcontentsline{toc}{section}{References}
\input{references}

\endgroup

\clearpage
\section*{Supplementary Information}
\addcontentsline{toc}{section}{Supplementary Information}
\setcounter{section}{0}
\setcounter{subsection}{0}
\setcounter{table}{0}
\setcounter{figure}{0}
\renewcommand{\thesection}{S\arabic{section}}
\renewcommand{\thetable}{S\arabic{table}}
\renewcommand{\thefigure}{S\arabic{figure}}
\input{supplement}
\end{document}

%% file: body.tex
\section*{Introduction}

Understanding when humans should work with generative artificial intelligence (AI) is increasingly important for individuals and organizations. Advances in AI have expanded the range of tasks these systems can perform, creating new choices about when to rely on human expertise, when to delegate work to AI, and when to combine the two \citep{fugener2026,raisch2025,eloundou2024}. A growing body of experimental research has examined these choices across different AI systems, tasks, and collaborative processes \citep{maier2026,dellacqua2026a,ju2025,doshi2024,goh2024}. The findings, however, are mixed: some studies show that working with AI improves performance, while others find that it can lead to worse outcomes \citep{maier2026,dellacqua2026a,ju2025,doshi2024,goh2024}. Despite this growing body of research, it remains unclear when human-AI combinations perform better than humans or AI working alone.

Meta-analyses can help address this question by combining results across experiments to estimate overall effects and examine how they vary across different conditions \citep{glass1976}. For example, \citet*{vaccaro2024} synthesized 370 unique effect sizes comparing the performance of human-AI combinations with that of humans and AI working alone. Their analysis identified human ability, AI ability, and task type as moderators of human-AI synergy and helped explain when combining the two improves performance.

\begin{figure}[htp]
\centering
\includegraphics[width=\linewidth]{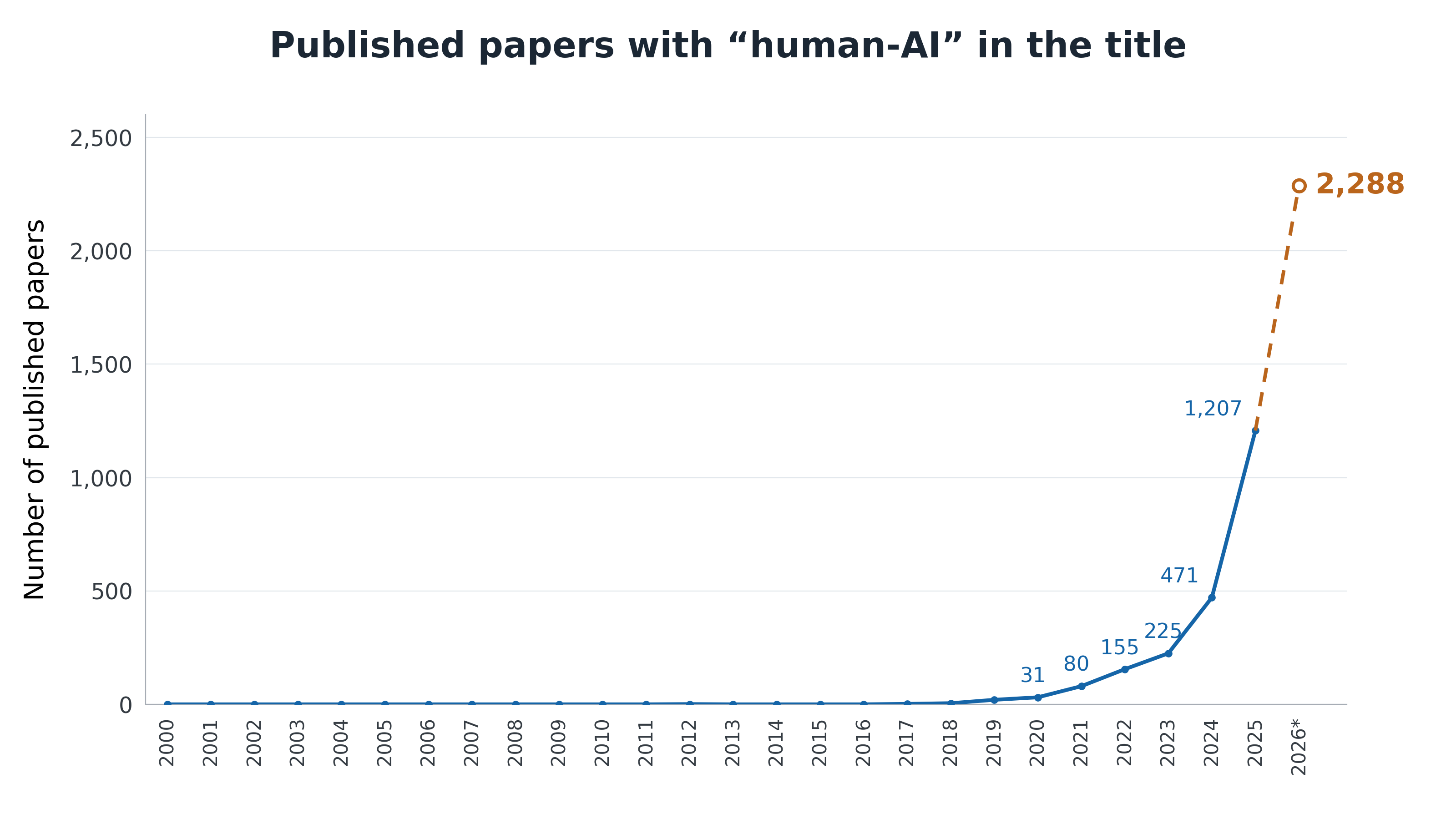}
\caption{\textbf{Published papers with ``human-AI'' in the title, 2000--2026.}}
\label{fig:publications}
\end{figure}

However, this type of conventional meta-analysis only reflects the evidence available at a particular point in time. Incorporating new studies involves additional searching, screening, reviewing, extracting, and analyzing, and the effort involved makes frequent updates difficult and, in practice, rare \citep{elliott2014}. As a result, the published and referenced synthesis often remains unchanged even as new findings emerge that may strengthen or challenge its conclusions. This limitation is particularly important for human-AI research, given both the growing volume of studies (see Fig.~\ref{fig:publications}) and the possibility that findings about earlier AI systems may not generalize to newer ones.

\begin{figure}[htp]
\centering
\includegraphics[width=.93\linewidth]{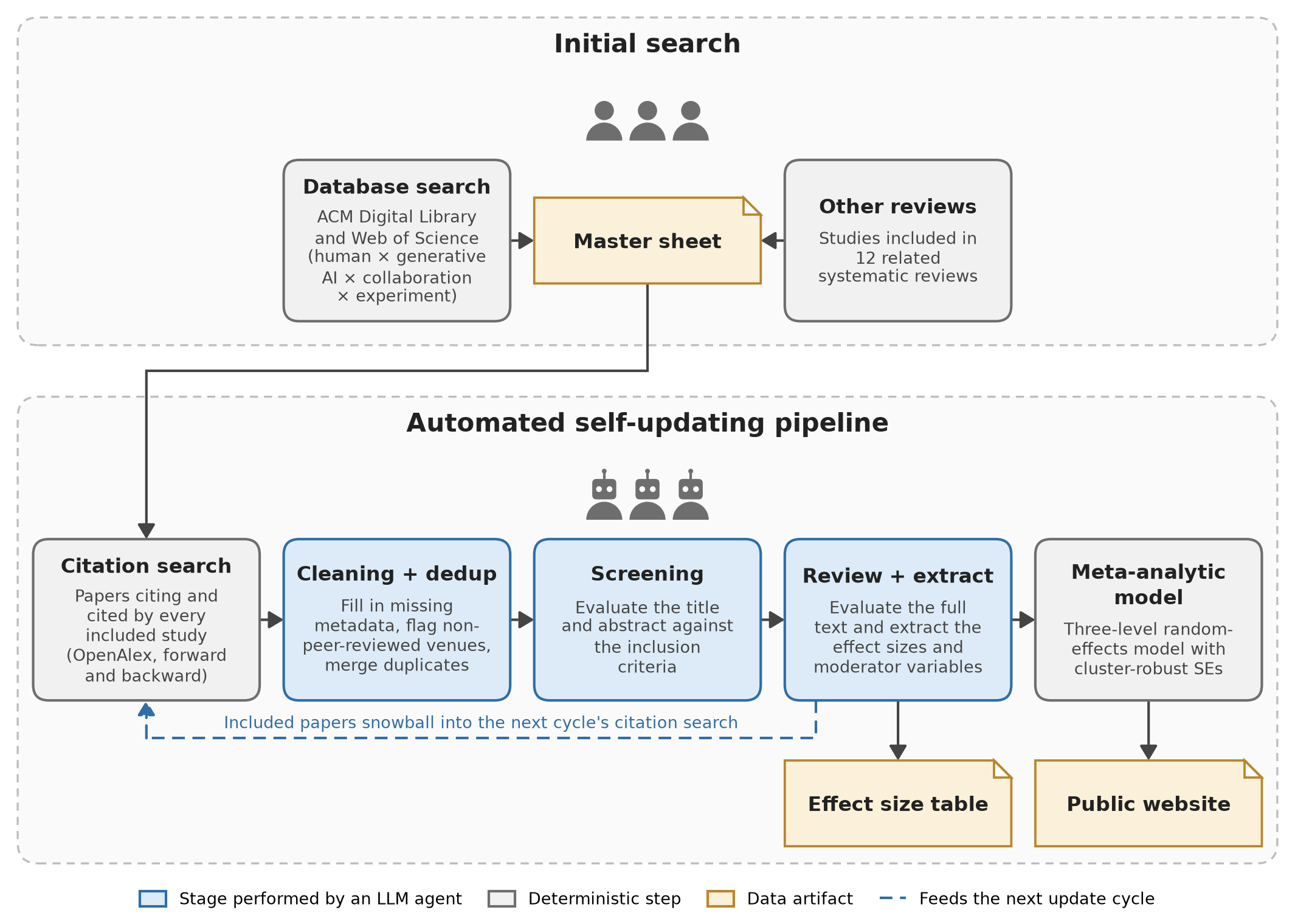}
\caption{\textbf{The AutoAnalysis pipeline.}}
\label{fig:pipeline}
\end{figure}

To address this challenge, we developed AutoAnalysis: an automated, self-updating meta-analysis that evaluates the performance of humans, AI, and human-AI combinations. Following a preregistered protocol, we designed a pipeline in which frontier AI agents perform each stage of a systematic review: identifying articles, screening abstracts, reviewing papers, extracting effect sizes and moderator variables, and synthesizing the resulting effect sizes and moderator variables, all with minimal human involvement (see Fig.~\ref{fig:pipeline}). As new studies become available, the pipeline automatically evaluates them for inclusion and incorporates eligible evidence into the meta-analysis. In other words, AutoAnalysis allows new research to inform and refine conclusions about when human-AI collaboration is useful.

AutoAnalysis' most recent run synthesizes 1,955 effect sizes from 420 papers, and it highlights that, among these studies, human-AI combinations performed significantly better than humans alone ($g = 0.53$, 95\% CI [0.45, 0.61], $p < 0.001$). However, among experiments that also evaluated AI alone, human-AI combinations performed significantly worse than the better of humans or AI working alone ($g = -0.35$, 95\% CI [-0.52, -0.18], $p < 0.001$). These findings indicate that, on average, combining humans and AI improved human performance but did not result in human-AI synergy. This pattern is consistent with \citet*{vaccaro2024}, despite being based on an almost entirely non-overlapping and more recent set of studies.

Importantly, human-AI performance varied across the actors, tasks, and processes examined. Our moderator analyses showed greater levels of human augmentation when the AI alone was stronger than the human alone than when the human alone was stronger ($p = 0.007$). This advantage was concentrated in decision tasks, where the AI alone outperformed humans alone by a wider margin than in creation tasks, where the two performed similarly ($p < 0.001$).

We also found that the process type was associated with different levels of human augmentation and human-AI synergy. In particular, when there was some division of labor between humans and AI---for example, they handled different sub-tasks---we saw greater levels of human augmentation ($p = 0.005$) and human-AI synergy ($p = 0.031$). Additionally, when the AI provided an explanation, we also saw greater levels of human augmentation ($p = 0.002$).

As AutoAnalysis incorporates new studies, it will show whether this pattern persists as AI systems improve and as researchers learn how to combine humans and AI more effectively.

\section*{Results}

\subsection*{Human-AI Augmentation and Synergy}

\begin{figure}[t]
\centering
\begin{subfigure}[t]{0.49\textwidth}
  \centering
  \includegraphics[width=\linewidth]{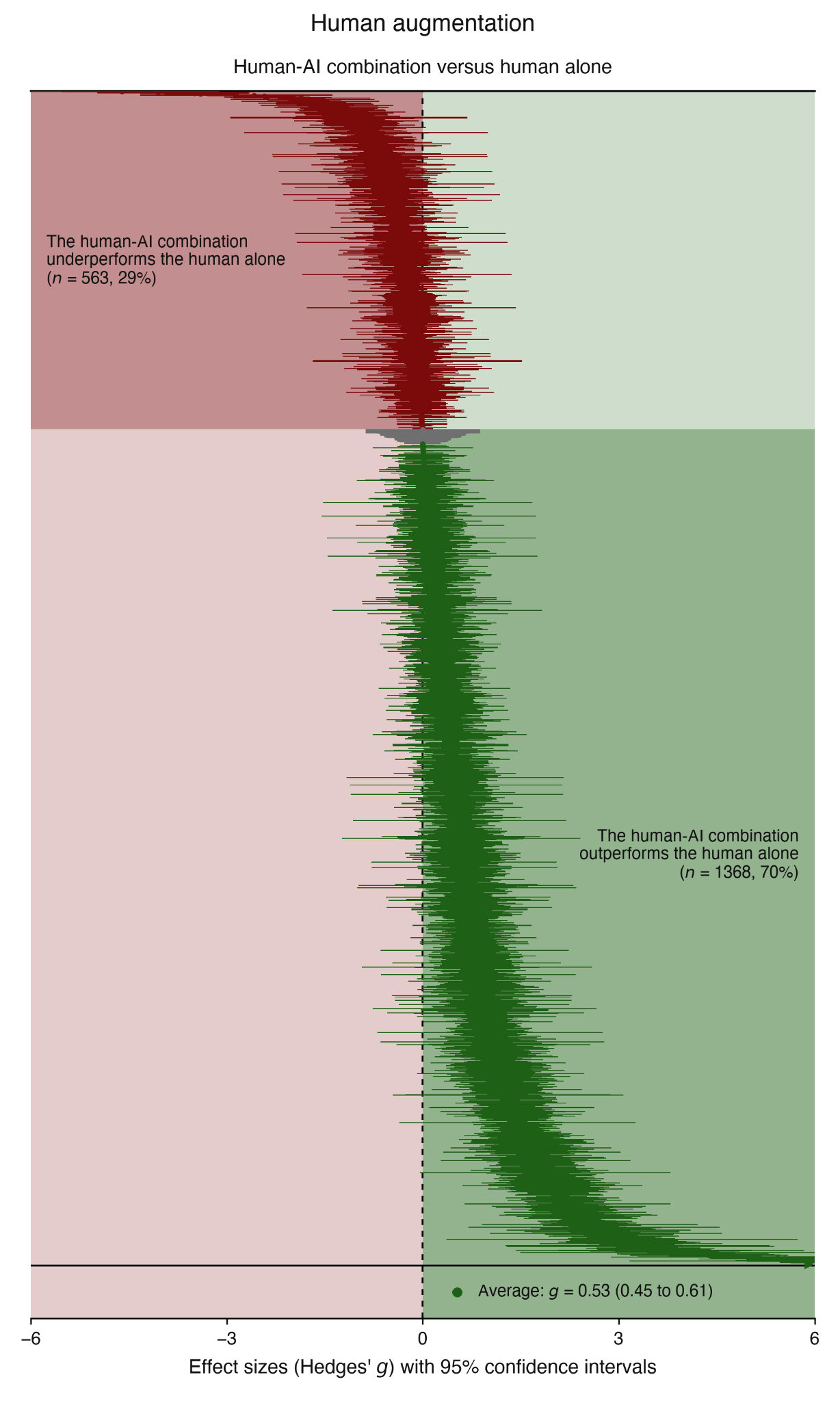}
  \label{fig:augmentation}
\end{subfigure}\hfill
\begin{subfigure}[t]{0.49\textwidth}
  \centering
  \includegraphics[width=\linewidth]{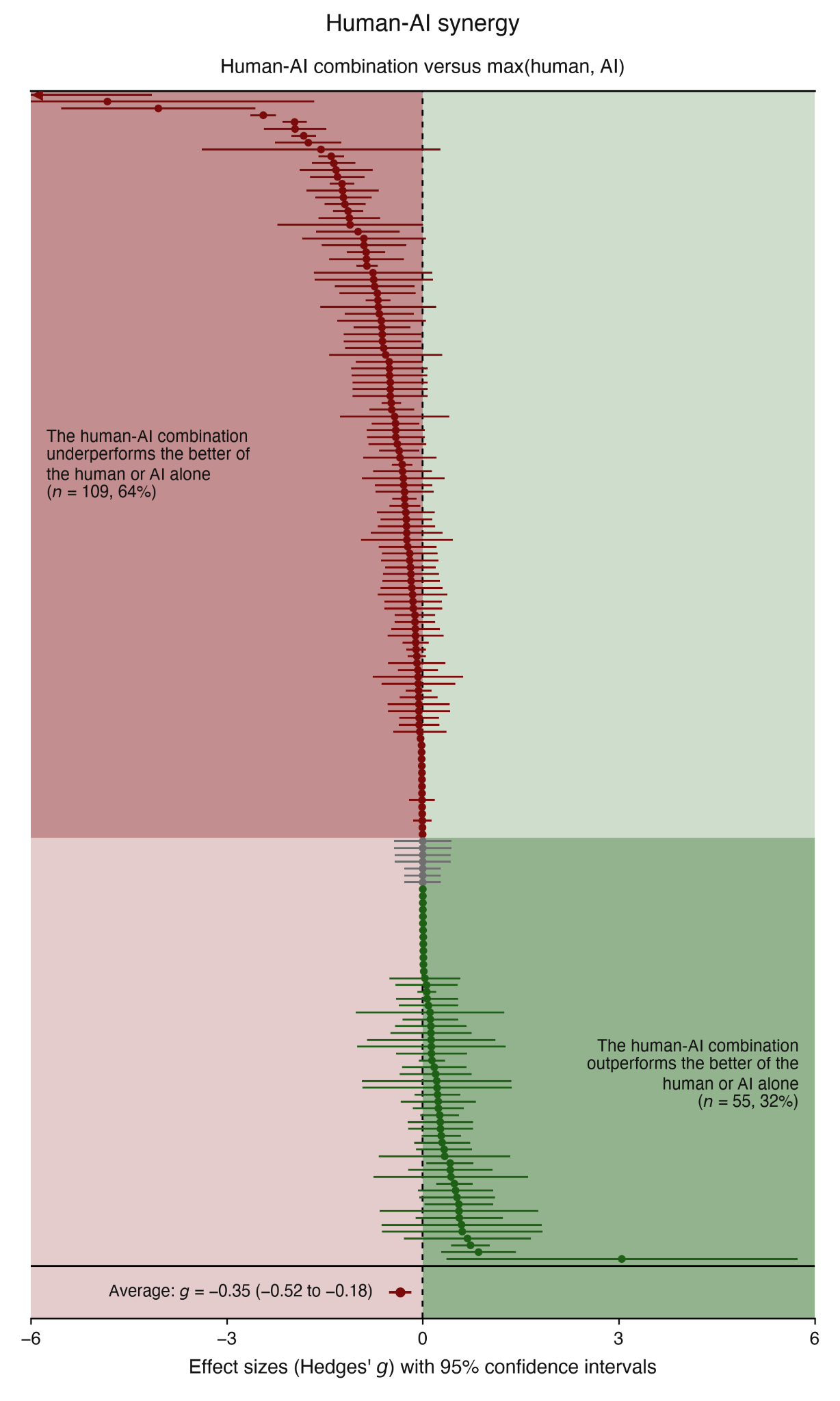}
  \label{fig:synergy}
\end{subfigure}
\caption{\textbf{Forest plots of human augmentation and human--AI synergy.}}
\label{fig:forest}
\end{figure}

On average, human-AI combinations performed substantially better than humans working alone (Fig.~\ref{fig:forest}). Across the 1,955 effect sizes, the pooled standardized mean difference was $g = 0.53$ (95\% confidence interval (CI) = [0.45, 0.61], $p < 0.001$). By conventional standards \citep{cohen1988}, this effect is considered ``medium-sized.'' Interestingly, it is of a similar magnitude to the $g = 0.64$ (95\% CI = [0.53, 0.74]) that we found in our 2024 meta-analysis, even though that paper involved mostly traditional (not generative) AI \citep{vaccaro2024}.

Although human-AI combinations outperformed humans alone, they did not, on average, achieve synergy (Fig.~\ref{fig:forest}). In fact, they underperformed either humans or AI working alone. Across the 171 effect sizes that also evaluated the AI system working alone, the difference between the human-AI system and whichever of the human alone or AI alone performed better was negative ($g = -0.35$; 95\% CI = [-0.52 to -0.18], $p < 0.001$). We thus find evidence of performance losses rather than synergy, replicating the pattern we reported for earlier human-AI systems ($g = -0.23$ in 2024) \citep{vaccaro2024}.

For both human-AI augmentation and human-AI synergy, we found high heterogeneity ($I^2$ = 92.9\% and 93.8\%, respectively). The moderator analysis below evaluates features of the tasks, the participants, the process and the study designs that can help account for this variation.

\subsection*{Moderators of Augmentation and Synergy}

The strongest predictor of whether AI helped humans perform better on a task was who performed better alone on that task: humans or AI. In the experiments that also measured the AI alone, human-AI combinations outperformed humans alone by a wide margin when the AI alone was stronger, but not when the human alone was stronger ($p = 0.007$). This gap between humans and AI was itself concentrated in decision tasks, in which the AI alone significantly outperformed humans alone ($p < 0.001$), whereas on creation tasks the two performed more alike.

We also found that the process type was associated with different levels of human augmentation and human-AI synergy. In particular, when there was some division of labor between humans and AI---for example, they handled different sub-tasks---we saw greater levels of human augmentation ($p = 0.005$) and human-AI synergy ($p = 0.031$).

Additionally, when the AI provided an explanation, we also saw greater levels of human augmentation ($p = 0.002$). Crowdworkers tended to benefit less from AI than non-crowdworkers ($p < 0.001$), and the task outputs ($p < 0.001$) and data ($p = 0.003$) were also associated with different levels of human augmentation.

\begin{figure}[htp]
\centering
\includegraphics[width=\linewidth]{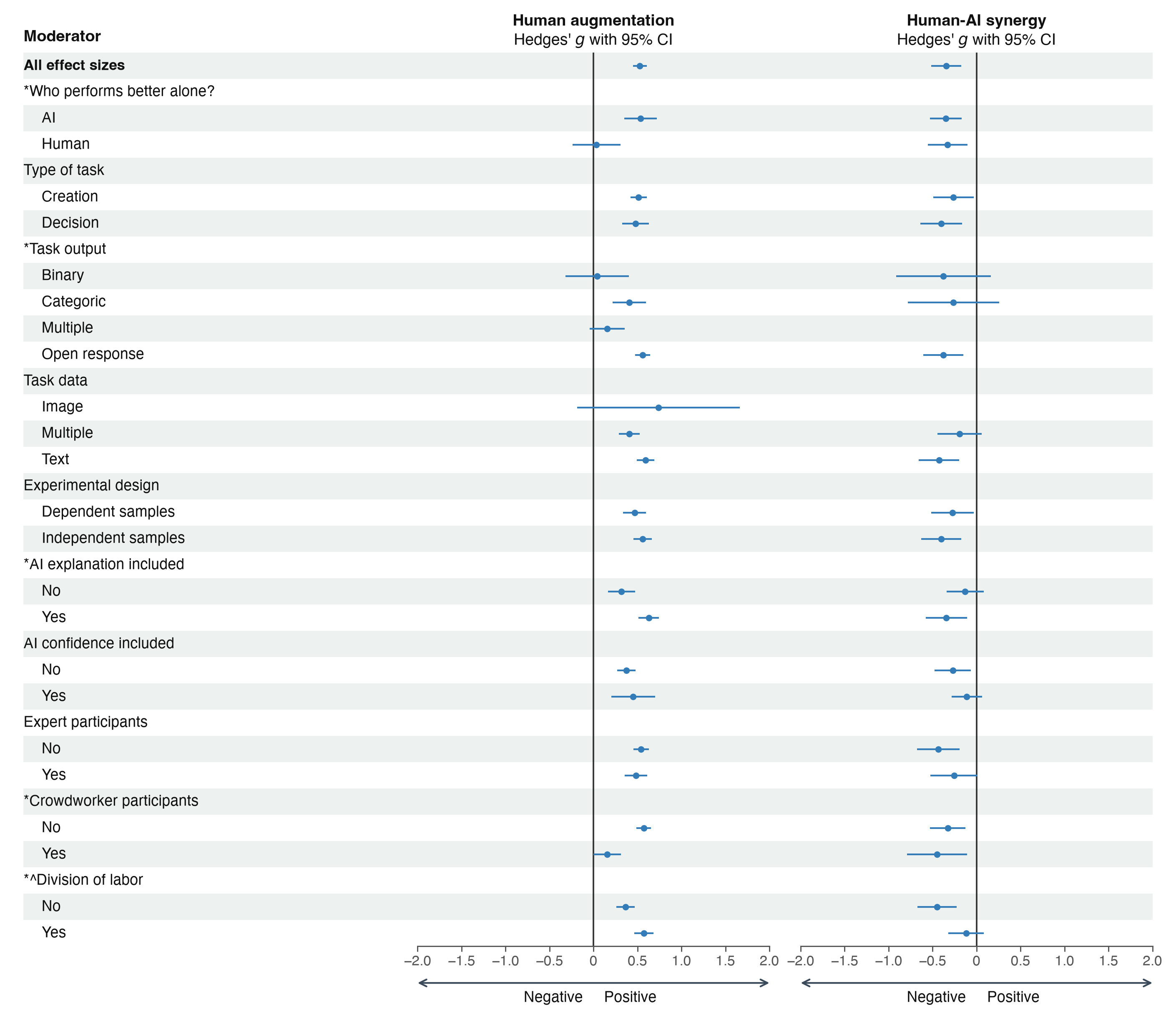}
\caption{\textbf{Moderator analyses of human augmentation and human--AI synergy.}}
\label{fig:moderators}
\end{figure}

\section*{Discussion}

Here, we introduce AutoAnalysis: an automated, self-updating approach to synthesizing evidence about human-AI collaboration. By using AI agents to identify studies, assess their eligibility, extract findings and update meta-analytic estimates, the platform enables a systematic review to evolve as new research becomes available. This capability is particularly valuable for human-AI research, where both the volume of evidence and the systems being studied are changing rapidly.

The evidence synthesized so far highlights an important gap between human augmentation and human-AI synergy. On average, human-AI combinations performed significantly better than humans working alone. However, among experiments that also evaluated AI alone, the combinations performed significantly worse than the better of humans or AI working alone. These results indicate that generative AI can improve human performance while leaving substantial opportunities to improve how human and AI capabilities are combined.

Interestingly, this pattern is consistent with \citet*{vaccaro2024}, whose meta-analysis examined largely non-generative AI systems. Despite drawing on an almost entirely non-overlapping body of more recent studies, we again found evidence of human augmentation alongside that of negative human-AI synergy. This consistency suggests that achieving synergy remains a challenge even with the broader capabilities of generative AI. It also highlights the need to complement advances in AI models with innovations in how humans and AI collaborate \citep{heyman2024,hosanagar2025,boussioux2024}.

Our exploratory moderator findings also suggest that such innovations should account for relative capabilities and the organization of work. Human augmentation was larger when AI outperformed humans alone, and division of labor was associated with better collaborative outcomes. These patterns motivate research on how to allocate subtasks according to human and AI strengths and integrate their contributions effectively \citep{dellacqua2026b,li2026,gu2026}. Because the moderator analyses do not establish causal relationships, controlled experiments are needed to test these possibilities. As such evidence accumulates, AutoAnalysis can help assess whether proposed design principles hold across different tasks, participants and AI systems, and whether they remain useful as those systems change.

\subsection*{Limitations}

We want to highlight several limitations that qualify our findings. First, the analysis reflects the studies captured by our search strategy and eligibility criteria. We included English-language, peer-reviewed experiments that reported sufficient quantitative information to estimate performance differences. These requirements exclude relevant evidence in other languages, unpublished studies and research awaiting peer review. In addition, the funnel asymmetry observed for augmentation raises the possibility of small-study effects. The absence of imputed studies in trim-and-fill analyses cannot fully resolve that concern \citep{sterne2011,shi2019}.

Second, the substantial heterogeneity across experiments limits what the pooled estimates imply for any particular application \citep{riley2011}. The studies examined different participants, AI systems, tasks and collaborative processes, and measured performance in different ways. Standardizing these outcomes allows us to synthesize the evidence, but does not make gains in speed, accuracy and creative quality equivalent for practical decisions. The estimates also do not establish how collaboration affects longer-term outcomes such as learning, skill retention or organizational performance \citep{bastani2025,brynjolfsson2025}. Moreover, our results about human-AI synergy depend on the relatively small subset of experiments that evaluated AI alone. These experiments may differ systematically from those that measured only human augmentation, limiting the generalizability of the synergy findings.

Third, the moderator analyses are exploratory, as characteristics of actors, tasks and processes overlap across studies, and separate meta-regressions cannot isolate their independent contributions. For example, an association between a process and better performance may reflect other features of the experiments using that process. As such, we view these analyses as useful information for identifying hypotheses for future human-AI experiments.

Finally, the automated pipeline introduces its own potential errors. Agents can miss eligible studies, misinterpret experimental designs or extract incorrect statistics, and errors at early stages can affect all subsequent analyses. Our human audits were small and conducted by the authors, and agreement with a separate audit agent does not establish accuracy against an independent reference standard. Models from different families may still make similar errors. Larger independent audits are therefore needed to estimate performance at each stage, including missed inclusions and errors in effect sizes and moderator coding. More validation will be particularly important as AutoAnalysis continues to ingest and evaluate new papers.

\subsection*{Conclusion}

AutoAnalysis provides an automated, self-updating synthesis of how humans and AI perform together across a growing body of experiments. The evidence synthesized so far shows that human-AI combinations improve on human performance, on average, but fall short of the better standalone performer in experiments that also evaluate AI alone. These findings highlight the importance of complementing advances in AI capabilities with innovations in how humans and AI collaborate. As new studies emerge, AutoAnalysis can track whether this pattern persists, refine our understanding of the conditions that support synergy and identify questions for further experimentation. By allowing evidence to accumulate alongside changes in AI technology and collaborative practices, the platform can help guide decisions about when to rely on humans, when to delegate to AI and how to combine their capabilities effectively.

\section*{Methods}

We conducted this meta-analysis following the standards suggested by Preferred Reporting Items for Systematic Reviews and Meta-Analyses (PRISMA) \citep{page2021}, and we preregistered the eligibility criteria, search strategy, coding scheme and primary analyses before running the pipeline (see here). Unlike a conventional meta-analysis, the review was carried out by AutoAnalysis, a pipeline in which frontier AI agents perform the search, screening, full-text review, data extraction and synthesis, and which re-runs on a schedule so that new evidence is incorporated as it is published. The methods below describe the pipeline as it was configured for the analyses in this paper, and the full specification (i.e., models, settings, and validation protocol) are provided in the Supplementary Information (see Sections~\ref{sec:architecture} and~\ref{sec:validation}). Note that all counts refer to the data release of October 26, 2026 that accompanies this paper. Because the analysis is self-updating, the counts on the public dashboard will differ as it incorporates papers published after this date.

\subsection*{Literature Review}

Per our preregistration, we applied the following criteria to select studies. First, the paper needed to present an original experiment with human participants that evaluated a scenario in which a human and a generative AI system worked together to perform a task. As such, experiments that used a simple predictive model, a shallow classifier, a rule-based decision aid, or a ``Wizard-of-Oz'' set-up were excluded, as were literature reviews, surveys, meta-analyses, position papers and purely theoretical work. Second, the paper needed to report the performance of the human alone and of the human-AI system according to some quantitative measure. We excluded papers that measured only the subjective experience of working with the system without measuring task performance. Third, we required the paper to report the experimental design, the number of participants in each condition, and the standard deviation of the outcome in each condition, or enough information to calculate it from other quantities. Fourth, we required the paper to be published in a peer-reviewed venue, so we excluded preprints, working papers, theses and technical reports. Finally, we required the paper to be written in English.

Because the pipeline discovers new studies by following citation links from included studies, we first assembled an initial set of papers in June 2026. We searched the abstracts of the Association for Computing Machinery Digital Library and the Web of Science Core Collection with a string crossing four components: a human, a generative AI system, a collaboration mechanism, and an experimental evaluation (see Supplementary Table~\ref{tab:search} for the exact syntax for each database). We also collected the primary studies included in twelve recent systematic reviews and meta-analyses of adjacent literatures: generative AI in education, software engineering, creativity, clinical medicine, work productivity and human-AI co-creativity (see Supplementary Table~\ref{tab:reviews} for more details). The database exports and review-derived records were merged and deduplicated by DOI and normalized title, and they yielded 1,759 unique records.

From this seed, the pipeline performs a forward and a backward citation search from every paper that passes full-text review. Using the OpenAlex API, an agent retrieves every article that cites an included paper, and every paper that is cited by an included paper. Because the search runs at the start of every update cycle, papers included in one cycle become seeds in the next, so the corpus expands by snowballing. This process yielded 15,495 unique records, again as of October 26, 2026.

The screening agent evaluated the title and abstract of each of these records against the eligibility criteria and included 2,403 papers. For each of these, the reviewing agent located a readable copy of the paper, directly for ACM publications and by web search across open-access sources otherwise, read the full text, and assessed it against the complete eligibility criteria. It returned a decision, a reason, and a flag indicating whether it had read the article body. This step included 420 papers.

For each included paper, the extracting agent produces one record per unique combination of experiment, treatment condition and outcome measure. Here, experiments are defined as samples of different sets of participants. For each record the agent extracts the means and standard deviations of task performance for the human alone, the AI alone (where reported), and the human-AI combination. The agent also extracts the number of participants in each condition, whether the comparison is between independent or dependent samples, and whether larger values of the metric indicate better performance. Finally, the agent extracts a series of moderator variables such as task type, task domain, and collaboration process. This process resulted in 1,955 unique effect sizes.

To validate this pipeline, we performed both human audits and agent audits. In particular, we reviewed a sample of 30 papers at each of the steps in the pipeline: screening, reviewing, and extracting. These checks were small and done by the authors rather than independent raters, so we treat them as a sanity check rather than a formal reliability study. To scale the audit process, we also built a second agent from a different model family than the pipeline's models, and had this agent review a random sample of papers at each stage of the process, and we find high inter-rater reliability (see Supplementary Information Section~\ref{sec:validation} for more details).

\subsection*{Data Analysis}

We computed Hedges' $g$ to measure the effect of combining humans and AI on task performance. For human augmentation, g represents the standardized mean difference between the performance of the human-AI system and that of the human alone. For human-AI synergy, $g$ represents the standardized mean difference between the performance of the human-AI system and that of the better of the human alone and the AI alone in the same experiment. This comparison is available only for the subset of experiments that evaluated the AI alone. We also compared the human-AI system to the AI alone and the human alone against the AI alone as secondary outcomes. We chose Hedges' $g$ because it is unitless, allowing comparison across different metrics, and because it corrects for the upward bias in the raw standardized mean difference (Cohen's $d$) in small samples. For each outcome metric we noted whether higher or lower values indicate better performance and, for lower-is-better metrics such as completion time and error rate, multiplied the standardized differences by $-1$, so that a positive g always indicates that the first-named group performed better.

The experiments we analyzed varied considerably: they evaluated different tasks, recruited participants from different backgrounds, used different AI systems and employed different experimental designs. Since we expected substantial between-experiment heterogeneity in the true effect sizes, we used a random-effects model that accounts for variance in effect sizes that comes from both sampling error and true variability across experiments. Additionally, many of the experiments generated multiple dependent effect sizes: they could involve multiple treatment groups compared with one control group, or assess performance according to more than one measure. The more commonly used meta-analytic models assume independence of effect sizes, which makes them unsuitable for our analysis. We therefore adopted the three-level meta-analytic model used in \citet*{vaccaro2024}, in which effect sizes are nested within experiments, with variance components for the variance in the observed effect sizes (first level), the variance between effect sizes from the same experiment (second level) and the variance between experiments (third level), estimated by restricted maximum likelihood.

Furthermore, we used robust variance estimation methods to compute consistent estimates of the variance of our effect-size estimates and, relatedly, standard errors, confidence intervals and statistical tests, which account for the dependency of sampling errors from overlapping samples that occurred in experiments that compared multiple treatment groups to a single control group. We defined clusters at the level of the experiment. When evaluating the significance of our results, we applied the Knapp and Hartung adjustment and computed a test statistic, standard error, $p$ value and confidence interval on the t distribution with $m-p$ degrees of freedom, where m denotes the number of effect-size clusters (that is, the number of experiments) and p the number of coefficients in the model.

As robustness checks we report four further estimators for every comparison: (1) a two-level random-effects model with cluster-robust standard errors clustered at the paper rather than the experiment level, which treats all effect sizes from one paper as dependent; (2) a three-level model with paper- and effect-level variance components and model-based standard errors; (3) a classic fixed-effect model; and (4) the unweighted mean of the effect sizes. All of these models are available on the AutoAnalysis website.

\subsection*{Moderator Analysis}

To examine how human-AI performance varies across actors, tasks and processes, we ran a separate meta-regression for each moderator variable, using the same three-level model, clustering and $t$-based inference as the main analysis. In each regression the moderator entered as a set of level indicators without an intercept, so that each coefficient is the pooled effect size for that level, estimated under a common pair of variance components, with cluster-robust standard errors by experiment and t inference on $m-p$ degrees of freedom. We tested whether the levels of a moderator differed with a robust Wald test of the $p-1$ contrasts against the first level, reported as an $F$ statistic on ($p-1$, $m-p$) degrees of freedom. We omitted effect sizes for which the moderator could not be determined from the paper and dropped levels with fewer than three effect sizes or fewer than two experiments, since a cluster-robust standard error is not defined for a level represented by a single experiment. Because the moderators overlap across experiments and we did not correct for the number of comparisons, we interpret differences between levels as descriptive rather than as evidence of statistically significant differences between categories.

\subsection*{Bias Tests}

In the context of our meta-analysis, publication bias may occur if researchers publish experiments that show evidence of human-AI synergy more frequently than those that do not. To evaluate this risk, we created funnel plots that graph the observed effect sizes on the $x$-axis and their standard errors on the y-axis and inspected them for asymmetry around the pooled estimate (see AutoAnalysis for variations of these funnel plots).

Second, we performed Egger's regression test, fitting the main three-level model with the standard error of each effect size entered as a moderator and testing its coefficient with cluster-robust standard errors on $m-2$ degrees of freedom. For human--AI synergy, Egger's test did not indicate significant asymmetry ($p = 0.098$), and trim-and-fill also did not impute any experiments. For our other outcome, human augmentation, Egger's regression did indicate funnel asymmetry ($p<0.001$), with smaller experiments reporting larger effects, but trim-and-fill imputed no missing experiments ($k = 0$), so the adjusted estimate equals the main estimate. Note that these results are as of the most recent run (October 26, 2026) and may change as AutoAnalysis ingests new papers and updates, accordingly. To this end, we provide self-updating versions of these plots on the AutoAnalysis webpage.

\subsection*{Sensitivity Analysis}

Following \citet{viechtbauer2010}, we also evaluated the robustness of our results to outlying and influential data points. Specifically, we identified outlier experiments, those in which effect sizes differed significantly from what the model fitted to the other experiments predicted ($p<0.05$). We also identified influential experiments, those which changed the pooled estimate by more than the conventional thresholds for Cook's distance, DFBETAS or leverage. For human-AI synergy, 8 of 54 experiments were flagged as outliers and none as influential. Excluding them attenuated but did not reverse the effect ($g = -0.20$ and $p = 0.004$, compared with $g = -0.35$ and $p< 0.001$). For human augmentation, 49 of the 451 experiments were flagged as outliers and none as influential. Excluding them gave a similar pooled effect ($g = 0.45$ and $p<0.001$) to the full data set ($g = 0.53$ and $p<0.001$).

%% file: references.tex
% Supplied bibliography, preserved as editable natbib-compatible entries.

%% file: supplement.tex
\section{AutoAnalysis architecture and configuration}\label{sec:architecture}

AutoAnalysis carries out every stage of a systematic review and meta-analysis---search, cleaning, screening, full-text review, data extraction, synthesis and reporting---and repeats the cycle on a schedule, so the published estimate tracks the literature as it grows.  Most stages are performed by a specialized agent built on a frontier language model. All agents read from and write to a single master record table with one row per candidate paper and that paper's complete decision history: where the record came from and when it was added, the screening decision and its reason, the inclusion decision and its reason, and whether extraction succeeded.

\begin{table}[htbp]
\centering\small
\begin{tabularx}{\linewidth}{@{}>{\raggedright\arraybackslash}p{0.18\linewidth} X >{\raggedright\arraybackslash}p{0.23\linewidth}@{}}
\toprule
\textbf{Stage} & \textbf{Task} & \textbf{Model}\\
\midrule
Cleaning & Check if a paper is peer-reviewed, and verify the title, authors, year, venue, DOI and abstract & gpt-6.1-sol (high) \\
\addlinespace[6pt]
Screening & Decide eligibility based on title and abstract & gpt-6.1-sol (high) \\
\addlinespace[6pt]
Reviewing and extracting & Locate a copy of the paper and, if found, make inclusion decision; if applicable, extract effect-sizes in the paper & gpt-6-astra (xhigh) \\
\addlinespace[6pt]
Auditing & Re-screen, re-review, or re-check a random sample of decisions & Claude Fable 5.1 (high) \\
\bottomrule
\end{tabularx}
\caption{\textbf{Stages of the AutoAnalysis pipeline.}}
\label{tab:stages}
\end{table}

\section{Initial citation search}\label{sec:initial-search}

We searched the abstracts of the ACM Digital Library and the Web of Science Core Collection in June 2026 with a string that captured our four inclusion criteria: (1) a human participant, (2) a generative AI system, (3) a collaboration mechanism, and (4) an experimental evaluation. Table~\ref{tab:search} provides the syntax for each database. We also collected the primary studies included in twelve recent systematic reviews and meta-analyses of adjacent literatures (Table~\ref{tab:reviews}).

\begin{table}[t]
\centering\small
\begin{tabularx}{\linewidth}{@{}>{\raggedright\arraybackslash}p{0.18\linewidth} X@{}}
\toprule \textbf{Database} & \textbf{Search string} \\ \midrule
ACM Digital Library & [Abstract: human] OR [Abstract: humans] OR [Abstract: participant] OR [Abstract: participants] AND [Abstract: "generative AI"] OR [Abstract: genAI] OR [Abstract: "large language model"] OR [Abstract: GPT] OR [Abstract: Claude] OR [Abstract: LLM] OR [Abstract: "diffusion model"] OR [Abstract: DALL-E] OR [Abstract: DALLE] OR [Abstract: Midjourney] OR [Abstract: "Stable Diffusion"] AND [Abstract: collaborate] OR [Abstract: assist] OR [Abstract: aid] OR [Abstract: interact] OR [Abstract: help] AND [Abstract: experiment] OR [Abstract: experiments] OR [Abstract: "user study"] OR [Abstract: "user studies"] \\
Web of Science Core Collection & AB=(human OR humans OR participant OR participants) AND AB=("generative AI" OR genAI OR "large language model" OR GPT OR Claude OR LLM OR "diffusion model" OR DALL-E OR DALLE OR Midjourney OR "Stable Diffusion") AND AB=(collaborate OR assist OR aid OR interact OR help) AND AB=(experiment OR experiments OR "user study" OR "user studies") \\
\bottomrule
\end{tabularx}
\caption{\textbf{Search strings for the database component of the seed dataset.}}\label{tab:search}
\end{table}

\begin{table}[t]
\centering\small
\begin{tabularx}{\linewidth}{@{}p{0.36\linewidth} X@{}}
\toprule \textbf{Review} & \textbf{Scope} \\ \midrule
\citet{alnaqbi2024} & Generative AI and work productivity across professional sectors \\[4pt]
\citet{belkina2025} & Case studies implementing generative AI in higher education \\[4pt]
\citet{deng2025} & Experimental studies of ChatGPT in education \\[4pt]
\citet{doo2026} & Experimental studies of ChatGPT and learning outcomes \\[4pt]
\citet{holzner2025} & Meta-analysis of generative AI and creativity \\[4pt]
\citet{delrio2025} & Generative AI, labor markets and worker productivity \\[4pt]
\citet{mo2026} & Meta-analysis of ChatGPT and undergraduate learning outcomes \\[4pt]
\citet{mohamed2026} & LLM assistants and software-developer productivity \\[4pt]
\citet{sergeyuk2026} & Human studies of AI programming assistants \\[4pt]
\citet{singh2026} & Human-AI co-creativity systems \\[4pt]
\citet{takita2025} & Diagnostic performance of generative AI versus physicians \\[4pt]
\citet{wang2026} & Human-LLM collaboration in clinical medicine \\[4pt]
\bottomrule
\end{tabularx}
\caption{\textbf{Systematic reviews and meta-analyses included in our initial search.}}\label{tab:reviews}
\end{table}

\section{Subsequent citation searches}\label{sec:citation-searches}

At the start of every update cycle, the search agent takes every paper that has passed the full-text review and queries OpenAlex for the works that cite it and the works it cites. New works are deduplicated against every record already in the dataset. Because papers included in one cycle become seeds in the next, the dataset expands outward from literature the pipeline has already included through this snowballing.

\section{Validation}\label{sec:validation}

\subsection{Human audits}\label{sec:human-audits}

As we developed and tested AutoAnalysis, we iteratively checked the agents' work at each stage of the pipeline. At screening we judged the title and abstract against the screening criteria, at full-text review we read the paper and judged it against the full eligibility criteria, and at extraction we extracted the effect sizes ourselves and compared our records with the agent's field by field. After we finalized the pipeline and ran the initial search, we reviewed another sample of 30 papers at each of these steps. These checks were small and done by the authors rather than independent raters, so we treat them as a sanity check rather than a formal reliability study.

\subsection{Agent audits}\label{sec:agent-audits}

We also built another agent to audit the decisions of other agents. It runs on Claude Fable 5.1 (high reasoning effort), a different model family from the pipeline's GPT models. At each of the three steps, it draws its own random sample of 30 papers and re-evaluates the other agents' decisions. For screening, the audit agent agreed with the screening agent on 90\% of the decisions, and the disagreements were borderline eligibility calls. For reviewing, the audit agent agreed with the reviewing agent on 83\% of the decisions, and the disagreements were cases in which the auditor could not access the full-text of the paper. For extracting, the audit agent reviewed the effect sizes from 30 papers field by field and found no incorrect mean, standard deviation or sample size. We plan to scale this audit process in the future.